\documentclass{article}
\usepackage[english]{babel}
\usepackage{footmisc}
\usepackage[letterpaper,top=2cm,bottom=2cm,left=3cm,right=3cm,marginparwidth=1.75cm]{geometry}
\usepackage{capt-of}

\usepackage[linesnumbered,ruled,vlined]{algorithm2e}

\SetKwInput{KwInput}{Input}                
\SetKwInput{KwOutput}{Output}              
\SetKwInput{KwDiff}{Differentiated Data}   
\SetKwInput{KwThres}{Threshold}            

\usepackage{amssymb}
\usepackage{amsmath}
\usepackage{graphicx}
\usepackage{natbib}
\usepackage{caption}
\usepackage{multicol}
\usepackage[colorlinks=true, allcolors=blue]{hyperref}
\usepackage{natbib}
\usepackage{xcolor}
\usepackage{placeins}
\usepackage{multirow}
\usepackage[textwidth=1.8cm, textsize=tiny]{todonotes}

\title{Deep Generators on Commodity Markets\\ Application to Deep Hedging}
\author{Nicolas BOURSIN\footnote{EDF Lab Singapore}, Carl REMLINGER \footnote{EDF Lab, FiME Laboratory, Université Gustave Eiffel }, Joseph MIKAEL\footnote{EDF Lab, FiME Laboratory}, Carol Anne HARGREAVES\footnote{National University of Singapore (NUS), Department of Statistics and Data Science}}
\date{}

\begin{document}
\maketitle

\begin{abstract}
\textit{Driven by the good results obtained in computer vision, deep generative methods for time series have been the subject of particular attention in recent years, particularly from the financial industry. In this article, we focus on commodity markets and test four state-of-the-art generative methods, namely Time Series Generative Adversarial Network (GAN) \cite{yoon2019time}, Causal Optimal Transport GAN \cite{xu2020cotgan}, Signature GAN \cite{ni2020conditional} and the conditional Euler generator \cite{remlinger2021conditional}, are adapted and tested on commodity time series.
A first series of experiments deals with the joint generation of historical time series on commodities.
A second set deals with deep hedging of commodity options trained on the generated time series. This use case illustrates a purely data-driven approach to risk hedging.
}
\end{abstract}

\section{Introduction}
Whether for pricing, market stress testing or market risk measurement, utilities are heavy users of time series models. For pricing, most of the models proposed in the literature focus on the design of stochastic models, among which we can cite the famous \cite{schwartz1997stochastic} or \cite{schwartz2000short}. We refer to \cite{deschatre2021survey} for a comprehensive survey of these stochastic commodity models. The design of these models is expensive, and once a model is available it remains to tackle the tedious task of its calibration. Also, a model usually cannot be updated quickly when market conditions change as this task requires in-depth expertise.\\ 

\noindent Two other arguments advocate even more for a change in the way commodity prices are simulated. 
First, the number and diversity of time series to be simulated increases with the emergence of renewable energies and new markets, making the model design even more complex.
Two other arguments militate even more for a change in the commodity price simulation paradigm.
First, the number and diversity of time series to be simulated increases with the emergence of renewable energies and new markets making the design of models even more complex.
Secondly, the need for a joint simulation of prices and volumes arises with the availability of new stochastic control tools based on machine learning and capable of manipulating a large class of models and in high dimension (\cite{fecamp2019risk} ).
The recent successive crises (sanitary, Texan, Russian), the consequences of which have been widely observed on the prices of raw materials, also plead for the rapid adaptation of models to new market conditions.\\ 

\noindent 
The rise of deep generative methods and Generative Adversarial Methods (GAN) for (static) images \cite{goodfellow2014generative,kingma2013auto} raises hopes for purely data-driven time-series simulators that could be more flexible and realistic.
The literature on deep generative methods for time series has thus particularly benefited from the Generative methods community.
We have thus seen proposals for neural network architectures capturing temporal dependencies, including recurrent neural networks and WaveNets \cite{esteban2017real,mogren2016c, oord2016wavenet,clark2019adversarial,wiese2020quant}.
To help the generator capture temporal dependencies or conditional distributions, recent works propose to embed the series of interest into a latent space. For example,
\cite{yoon2019time} transform the time series in a supervised way on a lower dimensional space and on which a GAN is applied. This method learns simultaneously the GAN and the latent space and requires five neural networks to work.
\cite{ni2020conditional} uses theoretically grounded signature-based embedding to extract meaningful features from trajectories, avoiding the need to learn the series representation. Methods based on the SDE formulation of the sequences have also been introduced to help the generator (\cite{kidger2021neural}).
Another method consists in designing objective functions to be optimized which take into account the temporal structure, such as the conditional distribution between time steps \cite{xu2020cotgan}.
Finally, other proposals include both SDE and conditional loss to design a time series generator without the need for a second neural network as a \cite{remlinger2021conditional} discriminator.

\noindent Some attempts to apply generative methods to time series involved in commodity markets have already been proposed. For example, \cite{chen2018model, qiao2020renewable} offers GANs whose purpose is to help in the operation and planning of electrical systems and to produce synthetic load curves. In particular, \cite{chen2018model} conditioned the generator based on weather events, such as high wind days or intense ramp events, and the time of year.\\ 

\noindent In this article, we propose to test state-of-the-art methods and adapt these frameworks to commodity markets (electricity, gas, coal and fuel). Additionally, in order to provide an operational metric, we propose to (deep) hedge  commodity options (including call option, spread options and an example of proxy hedging).
\\\\

\textbf{Contributions:}
\begin{itemize}
   
    \item an in-depth numerical comparison of the performance of deep generative models is performed on commodity price dataset,
    \item an application to deep hedging how what can be expected when combining deep generators with new approaches to hedging.
\end{itemize}

\section{Generative methods for time series}\label{sec:generators}
\subsection{Deep generative models}
\noindent Four time series generators are considered. They differ by the way in which the new data are constructed: sometimes by learning the representations of the series, sometimes by the choice of the objective function.
In the comparison, the network architecture originally proposed by the paper is kept, only the parameters  (such as the number of hidden layers) are changed for fairness.

\paragraph{Time Series GAN (TSGAN):} \cite{yoon2019time}  stands out by its specific training combining both supervised and unsupervised approaches. An embedding space is jointly learned with a GAN model to better capture the temporal dynamics. The generator thus produces sequences on a latent representation which are then reconstructed on the initial data space. By optimizing with both supervised and adversarial objectives, the model takes advantage of the efficiency of GANs with a controllable learning approach.
    
\paragraph{Causal Optimal Transport GAN (COTGAN):} \cite{xu2020cotgan} is an adversarial generator adapting the adapted Wasserstein distance to continuous time processes (\cite{backhoff2020adapted}). The model extends the regularized approach of the Wasserstein distance of \cite{genevay2018learning} to Causal Optimal Transport by adding a penalization on the traditional cost function. The discriminant networks learns to penalize anticipating transports and then ensures the temporal causality constraint. This model is theoretical sound, easy to implement and demonstrates less bias in learning than other GANs \cite{yoon2019time,donahue2018adversarial}

\paragraph{Signature GAN (SIGGAN):} \cite{ni2020conditional} combines a novel conditional Auto-Regressive Feedforward Neural Network (AR-FNN) for the generator with signature embedding. AR-FNN is a dedicated network architecture to learn auto-regressive structure of the sequence and maps past real time series and noise into future synthetic values. Signature (\cite{chevyrev2016primer}) is a theoretically grounded representation of series which characterize uniquely any continuous functions by extracting path features. Unlike classical Wasserstein GAN (\cite{arjovsky2017towards}), in this approach there are no needs to optimizing the discriminator, as conditional signature loss is used as critic.

\paragraph{Conditional loss Euler Generator (CEGEN):}  \cite{remlinger2021conditional} relies on a SDE representation of the time series and minimizes a conditional distance between transition distributions of the real and generated sequences. The stochastic process formulation helps the generator to build the series, while the conditional loss ensures the fidelity of the generations. The authors provide theoretical bounds error on the estimation of Itô processes.
As SIGGAN, CEGEN does not rely on a GAN framework and requires the training of one single neural network (\cite{ni2020conditional}).

\subsection{Metrics description}\label{sec:metrics}

One well-known challenge in the time series generation community is the lack of shared evaluation metrics (see \cite{wang2021generative}).
Temporal dependencies increase the difficulty of the analysis (\cite{eckerli2021generative,gao2020generative,eckerli2021generative}.
At a minimum, we can provide a set of basic metrics to characterize the most basic performances of the generations.

As done in \cite{remlinger2021conditional}, we consider the following metrics:
\begin{itemize}
    \item 
    Marginal based metrics includes classical statistics (mean, 95\% and 5\% percentiles denoted respectively \textbf{avg}, \textbf{p95}, \textbf{p05}). We compute the Mean Squared Error (MSE) over time of these statistics. 
    These metrics ensure that the marginal distribution is accurate and quantify the quality of the overall time series envelop.
    \item The temporal dependencies are measured with the MSE between quadratic variations (\textbf{QVar}) of the real and synthetic time series. The QVar is computed as follow: QVar$\left((X_t)_{t\geq0}\right)=\sum_{t} |X_{t+1}-X_{t}|^2$).
    \item Correlation structure \textbf{Corr} is evaluated with the time average MSE between the terms of the covariance matrices of synthetic and reference sequences. 
  
\end{itemize}
In Section \ref{sec:deephedging}, we propose market-practice based metrics.

\section{Numerical experiments}

\FloatBarrier

\subsection{Joint simulation of forwards}
\label{sec:jointsim}
\paragraph{Data} We consider the prices of 3-month futures contracts (3 MAH) on the electricity, coal, gas and fuel markets from January 21, 2020 to December 31, 2020. We generate sequences of 30 successive dates. Thus, deep generators must produce sequences in 4 dimensions.

\paragraph{Preprocessing} Commodity market prices have characteristics such as jumps and heavy tails that can be difficult to reproduce with generators like CEGEN because the latter generator is designed to repoduce an Itô process.
In order to facilitate the learning, a filter is operated on the data as a preprocessing task described in (pseudo code \ref{alg:filtercode}). By doing so, the trends of the original time series are preserved while avoiding extreme jumps.

{\centering
\begin{minipage}{.5\linewidth}

\begin{algorithm}[H]
\DontPrintSemicolon
  
  \KwInput{Historical time series $X_t$ }
  \KwDiff{$X_t^{diff}=X_t-X_{t-1}$}
  \KwThres{$q^{95}$ (quantile $95\%$ of $X_t^f$)}
  
   \For{$i=0...len(X_t^{diff})-1$}
   {
   		\If{$X_t^{diff}[i]>q^{95}$}
        {
            gap=$X_t^{diff}[i]-q^{95}$\\
            $X_t^{diff}[i]=q^{95}$\\
            $X_t^{diff}[i+1]=X_t^{diff}[i+1]$ $+$ gap\\
        }
        \ElseIf{$X_t^{diff}[i]<-q^{95}$}
        {
    	    gap=$-X_t^{diff}[i]-q^{95}$\\
            $X_t^{diff}[i]=q^{95}$\\
            $X_t^{diff}[i+1]=X_t^{diff}[i+1]$ $-$ gap\\
        }
   }
   $X_t^f=Cumulative\_sum(X_t^{diff})+X_t[0]$\\
  \KwOutput{Filtered time series $X_t^f$}
\caption{Jump Filter}
\label{alg:filtercode}
\end{algorithm}
\end{minipage}
\par
}

\FloatBarrier

\begin{figure}[!h]
\begin{minipage}{0.42\textwidth}
    \includegraphics[width=\textwidth]{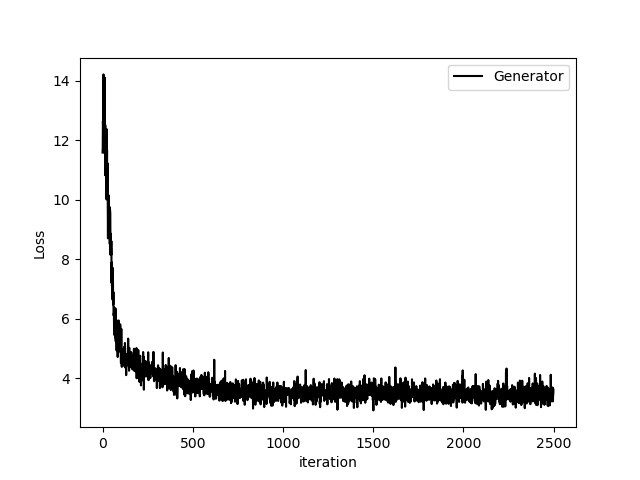}
    \caption*{CEGEN}
\end{minipage}
\hfill
\begin{minipage}{0.42\textwidth}
    \includegraphics[width=\textwidth]{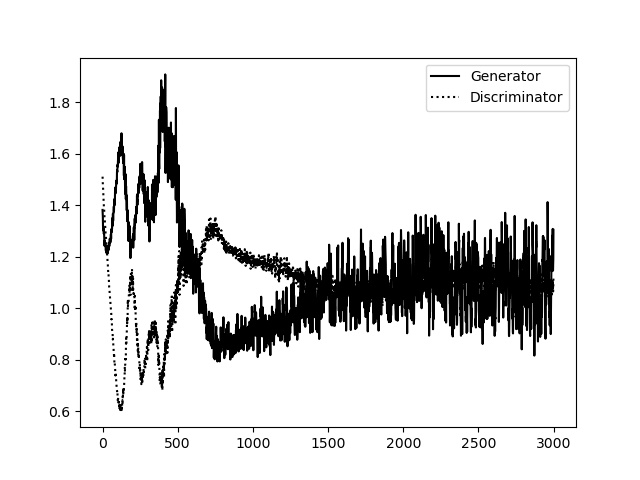}
    \caption*{TSGAN}
\end{minipage}
\begin{minipage}{0.42\textwidth}
    \includegraphics[width=\textwidth]{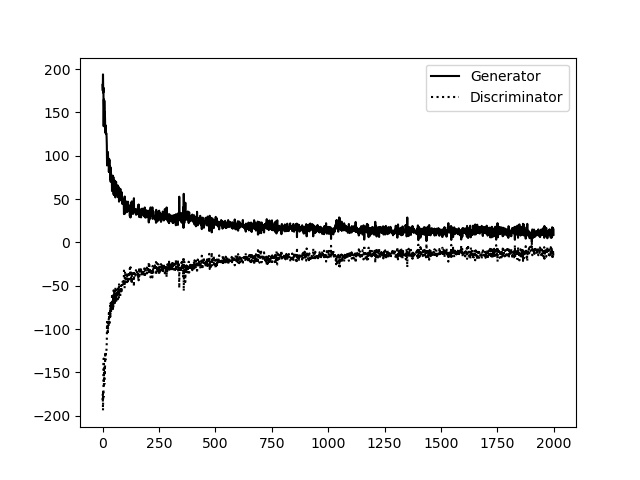}
    \caption*{COTGAN}
\end{minipage}
\hfill
\begin{minipage}{0.42\textwidth}
    \includegraphics[width=\textwidth]{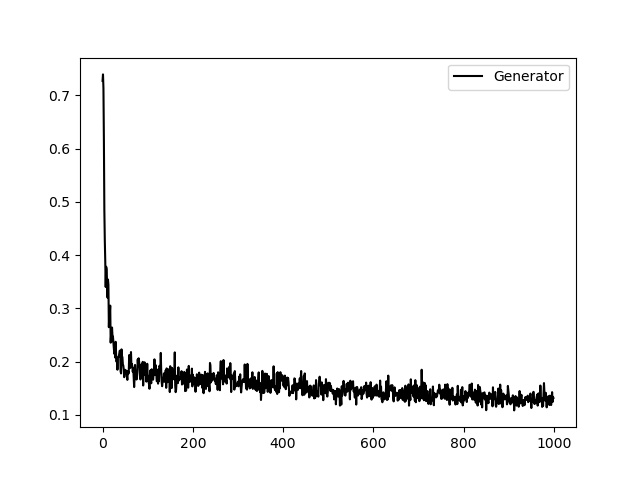}
    \caption*{SIGGAN}
\end{minipage}

\captionof{figure}{Generators losses curves. Dotted lines represent discriminator losses; full lines represent generator losses. For CEGEN and SIGGAN there is no discriminator neural network.}
\label{fig:genloss}
\end{figure}

\paragraph{Results}
As shown in figure \ref{fig:genloss}, all generators provide a satisfactory decrease of the loss function. The convergence of the GANs (here COTGAN and TSGAN) is subject to debate, but according to our experiences the number of iterations chosen gives correct and stable results that does not
depend on the initialization of the learning process.

\FloatBarrier

The table \ref{tab:joint} reports the performance of the four generators as well as the performance obtained by a 4-dimensional Geometric Brownian Motion (GBM) process. To calibrate the GBM volatility, we use the maximum likelihood estimator and we obtain $\hat{\sigma}=0.44$ for electricity, 0.50 for gas, 0.38 for oil and 0.25 coal. A reliable estimate of drift requires a lot of data. As we are simulating sequences of only 30 days it is not a determining factor, and we choose to set it to 0. Similarly, we report the correlations estimation between these products using the correlation matrix as follows:
$$\bordermatrix{
    & Elec. & Gas & Oil & Coal\cr
Elec. &  1 & 0.78  &  0.62  & 0.00\cr
Gas &  0.78   &  1  &  0.25 & 0.82\cr
Oil &  0.62  &  0.25 &  1  & 0.31\cr
Coal & 0.00  &  0.82  &  0.31  & 1
}$$

\begin{figure}[!ht]
\begin{minipage}{0.32\textwidth}
            \includegraphics[width=\textwidth]{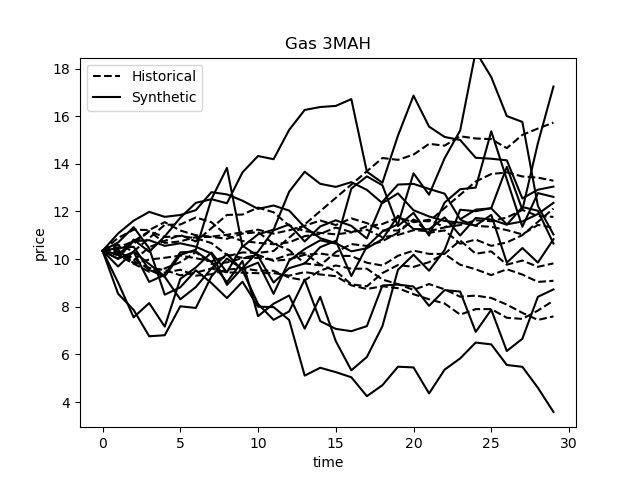}
              \caption*{GBM}
\end{minipage}
\begin{minipage}{0.32\textwidth}
    \includegraphics[width=\textwidth]{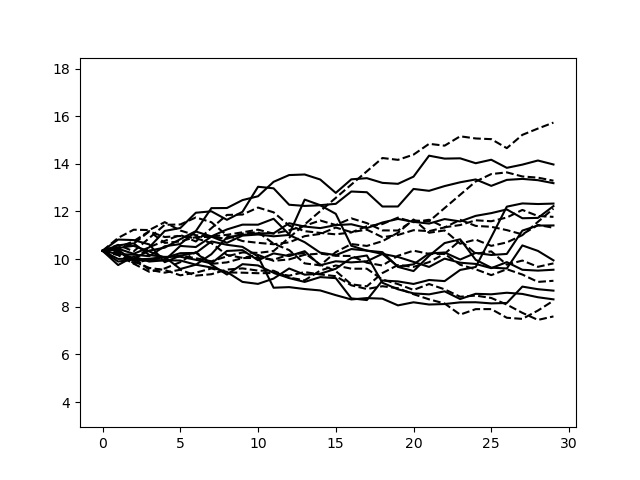}
    \caption*{CEGEN}
\end{minipage}
\begin{minipage}{0.33\textwidth}
    \includegraphics[width=\textwidth]{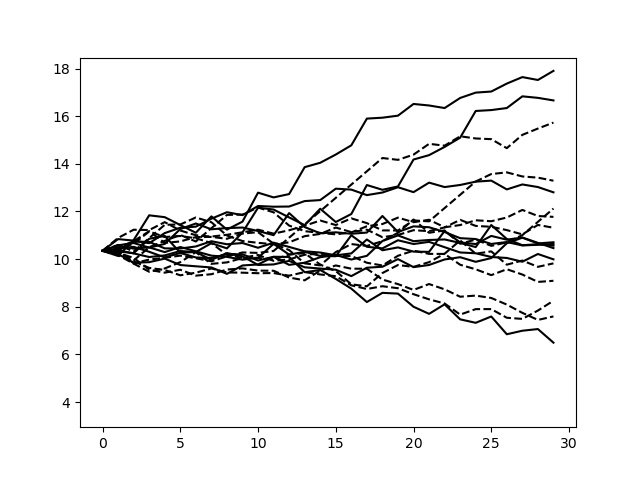}
    \caption*{TSGAN}
\end{minipage}
\begin{minipage}{0.33\textwidth}
    \includegraphics[width=\textwidth]{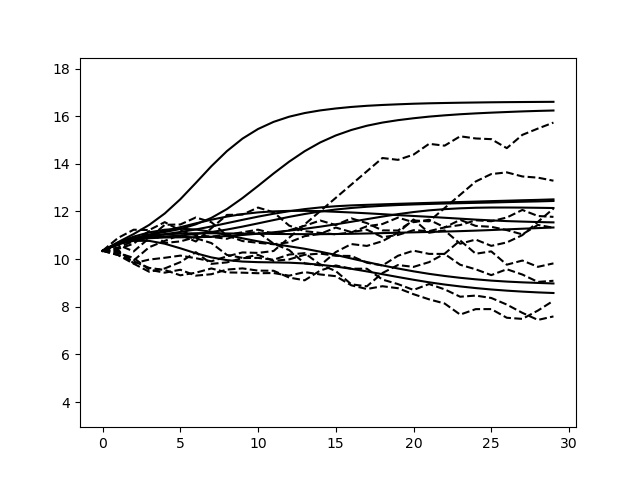}
    \caption*{COTGAN}
\end{minipage}
\begin{minipage}{0.32\textwidth}
    \includegraphics[width=\textwidth]{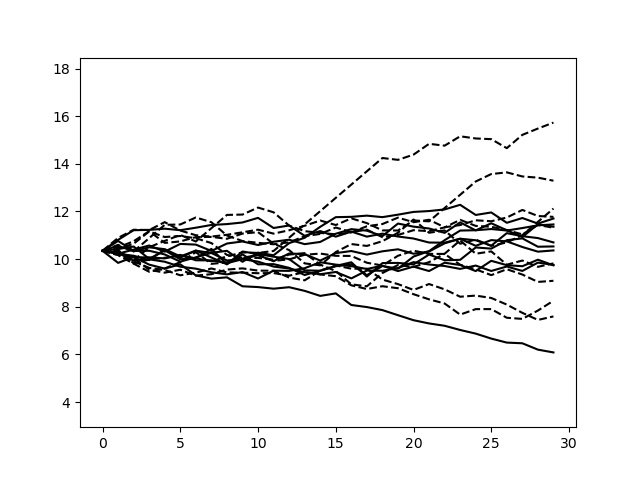}
    \caption*{SIGGAN}
\end{minipage}

\captionof{figure}{Generation of the 3MAH Gas. Dashed lines represent  trajectories of the historical time series; full lines represent samples generated by the generators.}
\label{fig:gasgraphes}
\end{figure}

\noindent The four generators estimate the marginal and conditional distributions better than the GBM. This indicates that the envelope estimate is better captured by these data-driven methods than by the GBM. Still on marginals,  CEGEN and TSGAN are relatively good in all markets: these two generators obtain 10 out of the 12 best metrics. It should be noted, however, that the TSGAN performs 100 times worse at the tail end of the process than near its centre of gravity (p95 and p05 compared to avg), which is not the case for CEGEN. On the temporal metric SIGGAN provides good QVAR which suggests that the time series signature is effective in representing the temporal features of the series. On the contrary, the Ito process does not seem to be a good representation of those features because CEGEN overestimates the QVAR. However, imposing this model allows to have consistent results on any dataset, and very reliable on the marginals. Similarly, the COTGAN adversarial loss performs well on the marginals but fails to replicate the temporal structure by flattening the time series as shown in Figure \ref{fig:gasgraphes}. TSGAN seems to be a good compromise as it represents well the marginal and transitional distributions, while remaining consistent on each dataset and being purely data-driven.

\begin{table}[!h]
\begin{center}
\begin{tabular}{|ll||ccc|c|}
\hline
\multicolumn{2}{|c||}{} & \multicolumn{3}{c|}{Marginal} & Temporal \\
&& p05 &     avg &    p95 &     qvar   \\\hline\hline
 &Elec. &7.27e-01 &2.90e-02 &4.60e-02 &7.49e+01\\
 GBM&Gas&6.60e-01 &5.85e-03 &1.25e-01 &8.90e+01\\
 &Oil&5.56e-01 &6.48e-02 &2.34e-01 &3.49e+01\\
 &Coal&5.70e-02 &2.91e-03 &7.16e-02 &4.13e+01\\\hline\hline
 &Elec.&\textbf{2.88e-03} &9.30e-04 &\textbf{3.31e-02} &2.17e+00\\
 CEGEN&Gas&\textbf{4.02e-03} &8.20e-04 &6.93e-02 &1.84e+00\\
 &Oil&2.09e-01 &3.63e-03 &3.10e-02 &4.60e+01\\
 &Coal&\textbf{1.97e-02} &\textbf{1.05e-03} &\textbf{2.95e-02} &8.84e+00\\\hline
 &Elec.&5.64e-02 &\textbf{2.60e-04} &5.70e-02& \textbf{1.51e-01}\\
 TSGAN&Gas&3.27e-02 &\textbf{1.70e-04} &\textbf{4.76e-02} &1.74e-01\\
 &Oil&4.34e-02 &\textbf{7.70e-04} &1.83e-01 &4.55e+00\\
 &Coal&1.84e-01 &5.87e-03 &3.03e-01 &7.32e+00\\\hline
 &Elec.&1.27e-01 &6.28e-02 &1.49e-01 &1.92e+00\\
 COTGAN&Gas&6.10e-02 &6.24e-02 &1.31e-01 &1.20e+00\\
 &Oil&4.01e-01 &6.63e-02 &\textbf{9.21e-03} &9.15e+00\\
 &Coal&1.60e-01 &1.70e-02 &3.65e-02 &1.51e+01\\\hline
 &Elec.&5.27e-02 &4.37e-03 &2.43e-01 &3.18e-01\\
 SIGGAN&Gas&3.77e-02 &2.89e-02 &5.06e-01 &\textbf{3.28e-02}\\
 &Oil&\textbf{1.68e-02} &1.03e-03 &1.05e-01 &\textbf{3.13e+00}\\
 &Coal&5.55e-02 &2.22e-02 &2.60e-01 &\textbf{3.24e+00}\\
\hline
\end{tabular}
\captionof{table}{Temporal and marginal metrics on joint simulation of electricity, natural gas, coal and oil 3 months forward prices. For all metrics, the lower, the better.}
\label{tab:joint}
\end{center}
\end{table}

\FloatBarrier

\FloatBarrier

\noindent As the \ref{tab:correlation} table shows, generators outperform GBM in terms of correlation. Surprisingly enough, CEGEN provides seven times better performance than the second best performing generator on this metric.

\begin{table}[!ht]
\centering
\begin{minipage}[b]{0.42\hsize}\centering

    \begin{tabular}{|l||c|}
\hline
 &Correlation \\\hline\hline
GBM&0.088 \\\hline\hline
CEGEN& \textbf{0.001}\\\hline
TSGAN&  0.025\\\hline
COTGAN& 0.007 \\\hline
SIGGAN& 0.051\\\hline

\hline
\end{tabular}
\captionof{table}{Correlation metric}
\label{tab:correlation} 
\end{minipage}
\hfill
\end{table}
\FloatBarrier

\paragraph{Conclusion} At this stage, CEGEN and TSGAN seem to stand out. However the results show that from one metric to another the performance of the generators varies a lot. The choice of metrics influences the results of the comparison. Hence it is not clear which (if any) of these generators could be good enough to be applied operationally.
One more time, the best model depends on what you want to do with it. That is why
in the next sections, we go beyond statistical metrics and offer financial and operational metrics on practical applications.

\subsection{Hedging related tests}
Synthetic time series can be used in risk management, portfolio structuring and, last but not least, in the pricing of derivatives and the derivation of associated hedging strategies. In the latter use case, the same model with two different calibration samples can lead to two different pricing and different hedging policies.
In this section, we use samples of the previously tested generators to (deep) hedge  an option on commodity derivatives.
Four deep hedgers are trained on synthetic samples from the four generators.
By comparing replication errors based on historical data, this test goes beyond the statistics of previous sections as it compared different generators on a metric of interest to the industry.

\begin{figure}[h]
    \centering
    \includegraphics[width = 0.9\textwidth]{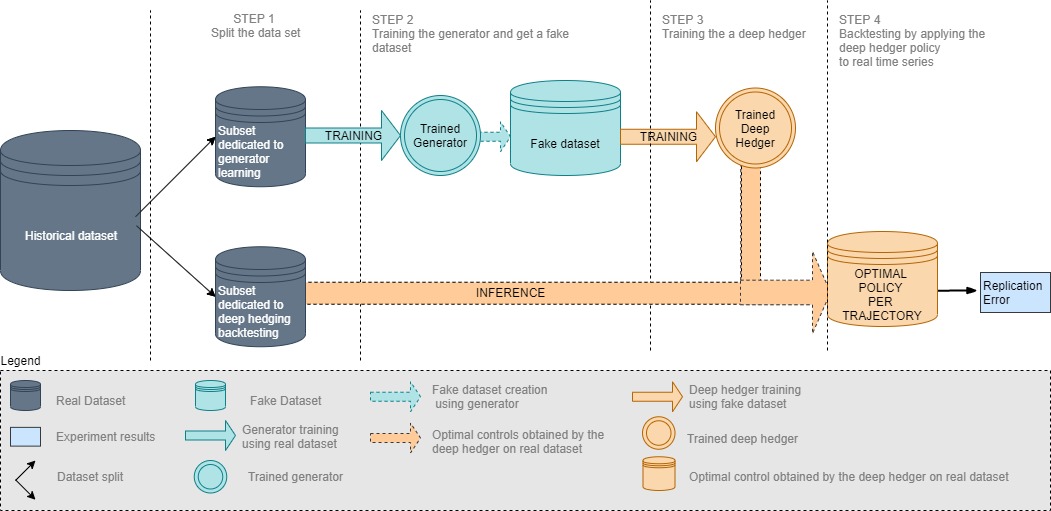}
    \captionof{figure}{Model-free deep hedger framework.}
    \label{fig:modelfreedeephedger}
\end{figure}

The global approach is illustrated in Figure \ref{fig:modelfreedeephedger}.
\subsubsection{Deep Hedging}
\label{sec:deephedging}
We are given a financial market operating in continuous time: we begin with a probability space $(\Omega, \mathcal{F}, \mathbf{P})$, a time horizon $0<T<\infty$ and a filtration $\mathcal{F} = (\mathcal{F}_t),$ ${0\leq t\leq  T}$ representing the information available at time $t$. We consider $d + 1 $ assets $S = {S}^0, \ldots, {S}^d$ available for trade. We denote $S^j_{t}$, $j=1\ldots d$ (resp. $S_t$ the value of $S^j$ (resp. $S$) at time $t$. For the sake of simplicity, we suppose a zero interest rate. \\
\noindent
We consider the hedging problem of a contingent claim paying $g(S_T)$ at time $T$ where $S_T$ denotes the contingent claim underlying vector. 
We consider a finite set of hedging dates $t_0<t_1<\ldots<{t_{N-1}} < {t_{N}} = T.$ 
A self-financing portfolio is a $d$-dimensional $(\mathcal{F}_t)$-adapted process $\Delta_t$.
Its terminal value at time $T$ is denoted $X_T^{\Delta,p}$ and satisfies:

\begin{eqnarray*}
    X_T^\Delta &=& p +\sum_{i=1}^d \sum_{j=0}^{N-1} \Delta^i_{t_j}(F^i_{t_{j+1}} -  F^i_{t_{j}}),
\end{eqnarray*}
where $p\in \mathbf{R}$ will be referred to as the premium. 
We search for an optimal strategy verifying:
\begin{equation}
 (p^{Opt}, \Delta^{Opt}) = Argmin_{p,\Delta}{\mathbf{E}\left[(X^\Delta_T- g(S_T))^2 \right]}. \label{eq:description}.
\end{equation}

\noindent To solve this problem, we favor the global approach described in \cite{fecamp2019risk} due to its ease of use. The optimal policy $\Delta$ is approximated by a feed-forward neural network called \textit{deep hedger} parameterized by a neural network. The training procedure consist in learning both the optimal controls and the premium.\\ 
A deep hedger optimal policy depends naturally on the simulations it's fed with during the learning procedure. In the following we consider as many deep hedger as deep generators, each model being trained on the simulations of one deep generator. At the end, we obtain four different hedging policies that we test on real price time series coming from historical datasets. It is worth denoting here that the deep hedger is an approximation that
contributes to the replication error of the hedged portfolios. In order to dissociate the very nature of the replication error that may come either from the underlying model or from the hedging policy estimation, we propose also to deep hedge the four dimensionnal GBM as benchmark instead of using the classical Black-Scholes theoretical hedging control. 
To evaluate the accuracy of our generators, we compare the replication error of the hedged portfolio (which is in other words the risk of the hedged portfolio). Moreover, if simulations are sufficiently realistic, we should obtain similar replication errors if we apply a hedging strategy on those simulations or on the historical time series. Because the test dataset is composed of historical time series, we do not expect a significant gap between the train and test loss curves of the deep hedgers.

\FloatBarrier
\subsubsection{Call Option Hedging with Deep Generative Methods}
\label{sec:deephedgingcalloption}
\paragraph{Test Case and Data} Deep hedgers are trained on synthetic data produced by the generative methods listed out in Section \ref{sec:generators}.
The hedging occurs on a daily basis on an at-the-money call option of payoff $g(S_T) = (S_T- K) ^+$. The corresponding values are $T=30/252$, $K=S_0$, $S_0^{elec}=41.41$ for electricity, $S_0^{gas}=10.34$ for gas, $S_0^{coal}=52.76$ for coal and $S_0^{fuel}=370.49$ for fuel.
At each learning iteration of the deep hedger, the generators simulate new data. The test set is composed of
211 actual historical price sequences (all available historical sequences). In addition to the
four deep generators and the deep hedged GBM, the results of $\Delta$ calculated from a theoretical Black-Scholes strategy are presented. 

\paragraph{Results} Table  \ref{tab:replloss} lists replication errors. The initial risk reports $\mathbb{E}[(g(S_T))^2]$ i.e. the initial risk carried by the option before its hedging. Better performance can be seen when the deep hedger is trained on SIGGAN, CEGEN and TSGAN time series rather than Geometric Brownian Motion (GBM). In this very specific case, despite the smooth COTGAN trajectories shown in Figure \ref{fig:gasgraphes}, the replication error of COTGAN remains small. This may not be the case for a payoff involving quadratic variations (like volatility options). Despite good hedging on gas and coal, SIGGAN performs poorly on oil and electricity which is not surprising as we noticed in table \ref{tab:joint} its inconsistency when simulating different data sets.
\begin{table*}[h!]
    \centering
    \begin{tabular}{|l||c|c|c|c|c|c|c|c|}
    \hline
     Repl. Loss  & INIT  & GBM&CEGEN & TSGAN & COTGAN & SIGGAN \\
        \hline\hline
Elec.  &6.13e+01 & 2.07e+00  & \textbf{8.66e-01}   & 1.27e+00 &1.65e+00&  3.22e+00 \\
Gas          & 4.91e+00& 1.42e-01 &3.32e-02& 2.99e-02&3.10e-01& \textbf{2.46e-02} \\
Oil          & 1.70e+03& 4.39e+02& 1.37e+02 & \textbf{9.91e+01}& 2.33e+02&4.54e+02\\
Coal         & 2.69e+01& 9.65e-01  & 3.67e-01   &4.29e-01  & 4.01e+00&\textbf{2.89e-01} \\
    \hline
    \end{tabular}
     
    \caption{Replication errors from deep hedgers trained on synthetic data of 3MAH forward prices (French market). The lower, the better.}
     \label{tab:replloss}
\end{table*}
 \FloatBarrier

\begin{figure}[h]
\begin{minipage}{0.32\textwidth}
    \includegraphics[width=\textwidth]{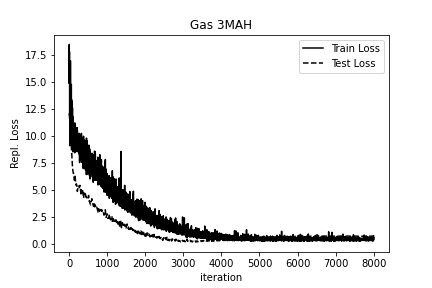}
    \caption*{Deep hedged GBM}
\end{minipage}
\begin{minipage}{0.32\textwidth}
    \includegraphics[width=\textwidth]{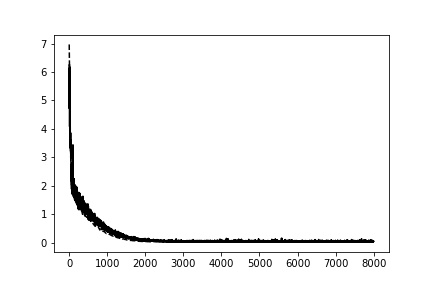}
    \caption*{CEGEN}
\end{minipage}
\begin{minipage}{0.33\textwidth}
    \includegraphics[width=\textwidth]{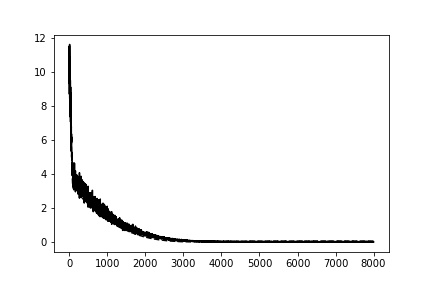}
    \caption*{TSGAN}
\end{minipage}
\begin{minipage}{0.33\textwidth}
    \includegraphics[width=\textwidth]{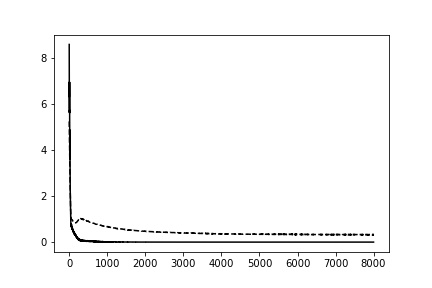}
    \caption*{COTGAN}
\end{minipage}
\begin{minipage}{0.32\textwidth}
    \includegraphics[width=\textwidth]{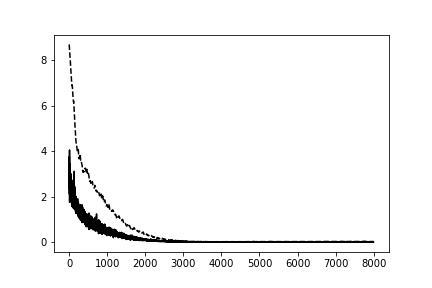}
    \caption*{SIGGAN}
\end{minipage}
\captionof{figure}{Losses curves of the deep hedgers training on the 3 MAH Gas call option. Dashed lines represent test losses computed on the historical time series; plain lines represent generator losses computed on synthetic time series.}
\label{fig:hedgerloss}
\end{figure}

\noindent Figure \ref{fig:hedgerloss} shows that the learning and testing loss curves of the deep hedgers are closer in the CEGEN and TSGAN cases than in the GBM, COTGAN, and SIGGAN cases. However the lower the difference between these two test curves the closer the simulations are to the real data. 

\FloatBarrier

\begin{figure}[h]
\begin{minipage}{0.32\textwidth}
            \includegraphics[width=\textwidth]{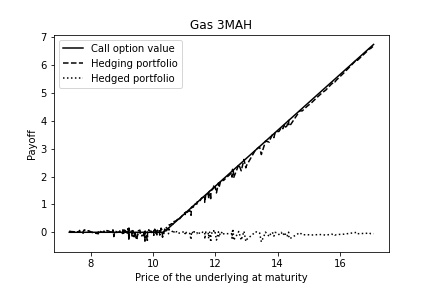}
              \caption*{Theoretical Black-Scholes}
\end{minipage}
\begin{minipage}{0.32\textwidth}
    \includegraphics[width=\textwidth]{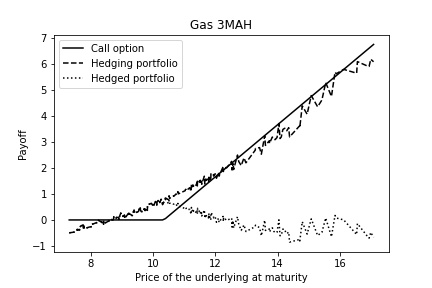}
    \caption*{Deep hedged GBM}
\end{minipage}
\begin{minipage}{0.32\textwidth}
    \includegraphics[width=\textwidth]{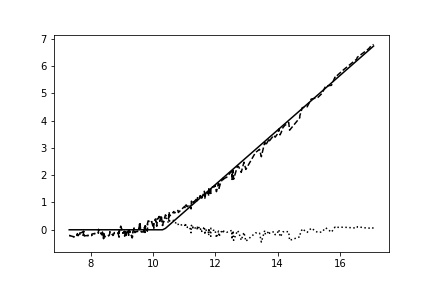}
    \caption*{CEGEN}
\end{minipage}
\begin{minipage}{0.33\textwidth}
    \includegraphics[width=\textwidth]{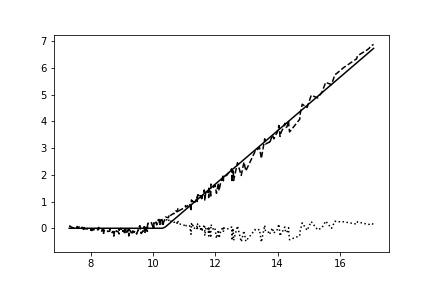}
    \caption*{TSGAN}
\end{minipage}
\begin{minipage}{0.33\textwidth}
    \includegraphics[width=\textwidth]{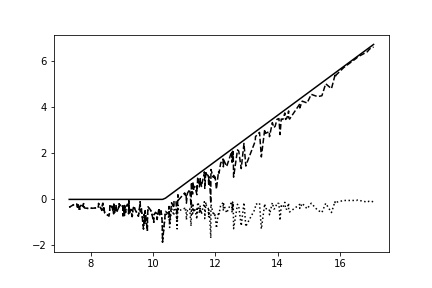}
    \caption*{COTGAN}
\end{minipage}
\begin{minipage}{0.33\textwidth}
    \includegraphics[width=\textwidth]{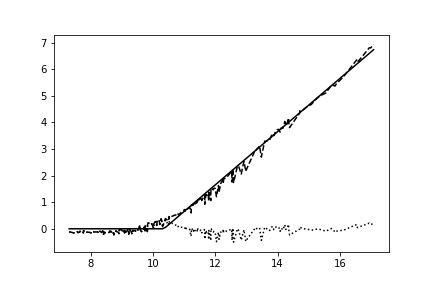}
    \caption*{SIGGAN}
\end{minipage}
\captionof{figure}{Dashed lines represent replicating portfolios values at maturity; full lines represent call option payoffs; dotted lines represent the hedged portfolio selling the call at time 0 and hedging its risk by trading the underlying asset.}
\label{fig:replgraph}
\end{figure}

\noindent 
Figure \ref{fig:replgraph} shows the value of the replicated portfolios at expiration as a function of $S_T$ and plot the corresponding call option payoffs. For the four generators, we find the expected broken-line shape of the payoff $g(x)=(x-K)^{+}$. In the COTGAN's case, by far the payoff seems to be respected in consistency with the results it obtains on the marginals. However, if we take a closer look, we see an inability to react sufficiently well to variations around the strike. In this situation, price volatility becomes prevalent and the smooth COTGAN simulations yield an approximate hedging. In the case of the deep-hedged GBM, the broken line shape is not well recovered as well, which is consistent with the poor results of the marginals in Table \ref{tab:joint}. We get good hedging strategies with the other three generators, though with some difficulties around the strike. 
\FloatBarrier

\begin{figure*}[!h]
\begin{minipage}[!ht]{0.40\hsize}
    \footnotesize
    \centering
    \begin{tabular}{|l||c|c|}
    \hline
    Repl. Loss   & INIT &Theoretical Black-Scholes\\
    \hline\hline
    Elec.  &6.13e+01 &    1.41e-01\\
    Gas          &  4.91e+00 &   1.00e-02\\
    Oil          & 1.70e+03 &  1.50e+01\\
    Coal         & 2.69e+01 &    5.21e-02\\
    \hline
    \end{tabular}
    \captionof{table}{Replication errors of the theoretical Black-Scholes strategy.}
\label{tab:repllossBS} 
\end{minipage}
\hfill
\begin{minipage}[!ht]{0.45\hsize}
    \centering
    \includegraphics[width=0.99\textwidth]{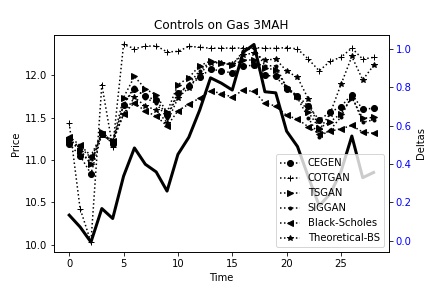}
    \footnotesize 
    \captionof{figure}{Controls proposed by each strategy. The plain line represents a 3MAH gas prices sequence. The other lines are the quantity of the underlying one needs to have at each time step to hedge their risk according to each strategy.}
    \label{fig:controls}
\end{minipage}
\hfill

\end{figure*}

\noindent Table \ref{tab:repllossBS} shows the replication error when using the theoretical Black-Scholes strategy and Figure \ref{fig:hedgerloss} shows the corresponding replication portfolios. The difference between these values and the deep hedged GBM comes from the approximation error of the deep hedger. Further work could be done to improve the deep hedger performances but at this step, it is still good enough to compare the accuracy of different generators. \\

\noindent If on this classical delta hedging case, the use of reinforcement learning seems useless compared to a simpler and more efficient Black-Scholes strategy, we will see in the following sections that reinforcement learning is advantageous because of its ability to hedge more exotic options and its flexibility when dealing with incompleteness (e.g. liquidity constraints).\\

\noindent Figure \ref{fig:controls} shows sample controls from the different hedging strategies. CEGEN, TSGAN, and SIGGAN appear to converge on the same policy while the COTGAN control is the furthest from these policies and advocates a policy that does not appear to respond well enough to price changes. This is consistent with the table \ref{tab:replloss} and the unrealistic smooth curves shown in Figure \ref{fig:gasgraphes}.

\FloatBarrier

\paragraph{Conclusion} When using a deep hedger, the generators appear to outperform the traditional GBM model. Again, TSGAN and CEGEN stand out from the other generators in terms of risk reduction. These results are consistent across commodities, which is not the case for SIGGAN. 
\FloatBarrier
\subsubsection{Option Hedging using a Proxy with Deep Generative Methods}
\paragraph{Test Case and Data} We use the same framework as in the section on deep hedging to hedge an at-the-money call option on gas by trading only coal, with a payoff $g(S_T) = (S_T^{gas}- K) ^+$, $K=S_T^{gas}=10.34$. The high correlation (close to $0.8$) between these two commodities justifies this test. The other data are the same as in section \ref{sec:deephedgingcalloption}.

\paragraph{Results} Table \ref{tab:repllossproxy} shows the replication errors of the proxy hedge. As expected, the risk reductions are smaller than in the previous section because we cannot directly trade the underlying asset on which the option is based.
\begin{table*}[h!]
    \centering
    \begin{tabular}{|l||c|c|c|c|c|c|}
    \hline
    Repl. Loss  & INIT  & GBM&CEGEN & TSGAN & COTGAN & SIGGAN \\
        \hline\hline
Gas (proxy: Coal)&4.91e+00&2.64e+00&\textbf{2.07e+00}&3.99e+00&6.61e+00&2.33e+00\\

    \hline
    \end{tabular}
     
    \caption{Replication errors from proxy deep hedgers trained on synthetic data on 3MAH forwards (French market). The lower, the better.}
     \label{tab:repllossproxy}
\end{table*}

\noindent Figure shows that CEGEN is the only strategy whose train and test losses overlap. This suggests that CEGEN provides a viable hedging strategy. 
  
  \begin{figure}[h]
\begin{minipage}{0.32\textwidth}
    \includegraphics[width=\textwidth]{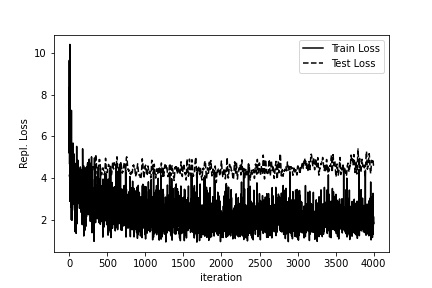}
    \caption*{Deep hedged Black-Scholes}
\end{minipage}
\begin{minipage}{0.32\textwidth}
    \includegraphics[width=\textwidth]{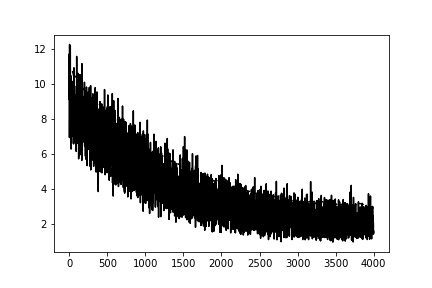}
    \caption*{CEGEN}
\end{minipage}
\begin{minipage}{0.33\textwidth}
    \includegraphics[width=\textwidth]{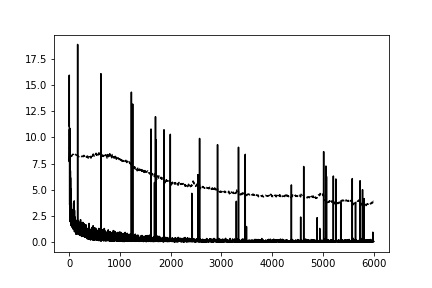}
    \caption*{TSGAN}
\end{minipage}
\begin{minipage}{0.33\textwidth}
    \includegraphics[width=\textwidth]{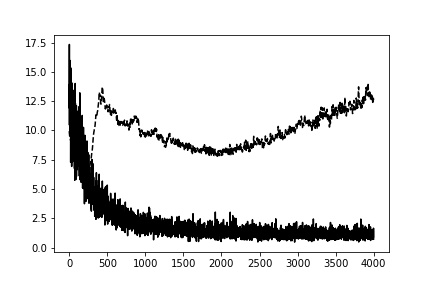}
    \caption*{COTGAN}
\end{minipage}
\begin{minipage}{0.32\textwidth}
    \includegraphics[width=\textwidth]{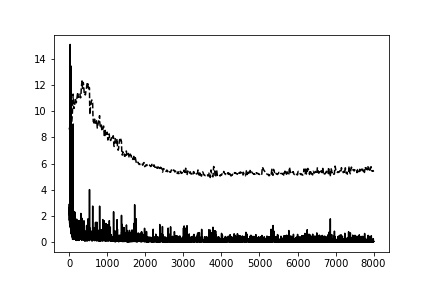}
    \caption*{SIGGAN}
\end{minipage}
\caption{Losses curves of the deep hedgers training on the 3MAH Gas and hedging on 3MAH Coal. }
\label{fig:hedgerlossproxy}
\end{figure}

\noindent  The key to successful proxy hedging is to correctly measure the correlations between commodities. CEGEN's performance underscores that one of its strength lies in its ability to replicate these correlations, according to the chart \ref{tab:correlation}. \\

\begin{figure}[h]

\begin{minipage}{0.32\textwidth}
    \includegraphics[width=\textwidth]{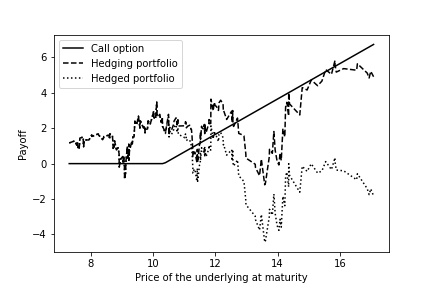}
    \caption*{Deep hedged Black-Scholes}
\end{minipage}
\begin{minipage}{0.32\textwidth}
    \includegraphics[width=\textwidth]{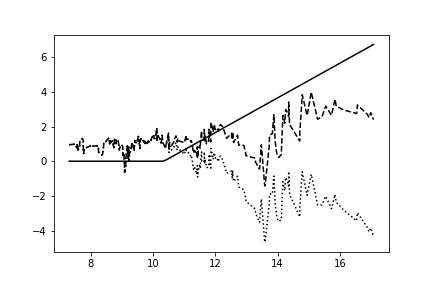}
    \caption*{CEGEN}
\end{minipage}
\begin{minipage}{0.33\textwidth}
    \includegraphics[width=\textwidth]{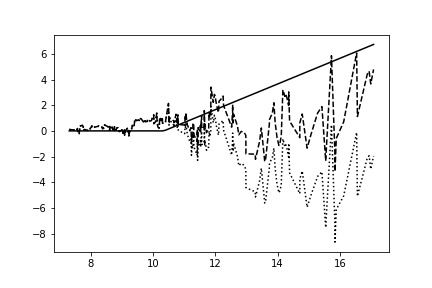}
    \caption*{TSGAN}
\end{minipage}
\begin{minipage}{0.33\textwidth}
    \includegraphics[width=\textwidth]{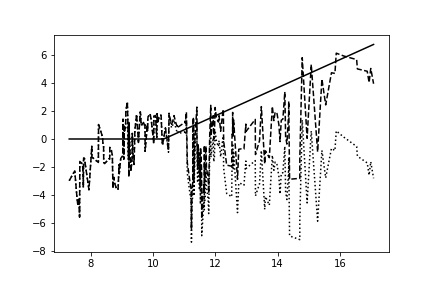}
    \caption*{COTGAN}
\end{minipage}
\begin{minipage}{0.33\textwidth}
    \includegraphics[width=\textwidth]{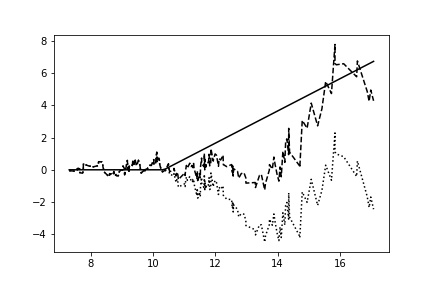}
    \caption*{SIGGAN}
\end{minipage}
\caption{Replicating portfolios of the proxy hedging call option.}
\label{fig:replgraphproxy}
\end{figure}

\noindent Figure \ref{fig:replgraphproxy} shows that the replication portfolios' values are not overlapping their corresponding option payoff as in the previous section, but the deep hedgers seem to be reactive after the strike level.

\FloatBarrier
\paragraph{Conclusion}  In this test case, the representation of the correlation is an essential asset to obtain good results. In this situation, CEGEN outperforms other generators, including the GBM model. 

\FloatBarrier

\subsubsection{Spread Option Hedging with Deep Generative Methods}

\paragraph{Test Case and Data} In this section, we hedge an at-the-money call option on gas and coal with a payoff $g(S_T) = (S_T^{coal}-S_T^{gas}- K) ^+$. The spread is the price difference between these two commodities. The strike is the initial spread ($K=S_0^{coal}-S_0^{gas}=42.41$). Unlike the previous section, we can trade gas and coal, the underlyings on which the option is based. Thus, while the success of the hedge still depends heavily on a correct representation of the correlation, it is no longer the only asset for learning a good strategy. The other data are the same as in Section \ref{sec:deephedgingcalloption}..

\paragraph{Results} Since a good replication model must combine the characteristics of the two previous sections, i.e. a good understanding of the distribution of the underlyings taken one by one (gas and coal in this case), and of their joint law, it is not surprising that CEGEN still performs well because it has good results on the first two use cases (see table \ref{tab:repllossspread} and \ref{tab:replloss})

\begin{table*}[h!]
    \centering
    \begin{tabular}{|l||c|c|c|c|c|c|}
    \hline
    Repl. Loss  & INIT  & GBM&CEGEN & TSGAN & COTGAN & SIGGAN \\
        \hline\hline
 Gas \& Coal spread &1.94e+01&3.70e-00&\textbf{3.03e-01}&8.35e+00&2.00e+00&1.37e+01\\
    \hline
    \end{tabular}
     
    \caption{Replication errors from spread option deep hedgers trained on synthetic data on 3MAH forwards (French market). The lower, the better.}
     \label{tab:repllossspread}
\end{table*}

\begin{figure}[h]
\begin{minipage}{0.32\textwidth}
    \includegraphics[width=\textwidth]{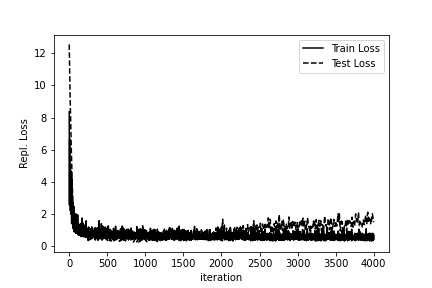}
    \caption*{Deep hedged Black-Scholes}
\end{minipage}
\begin{minipage}{0.32\textwidth}
    \includegraphics[width=\textwidth]{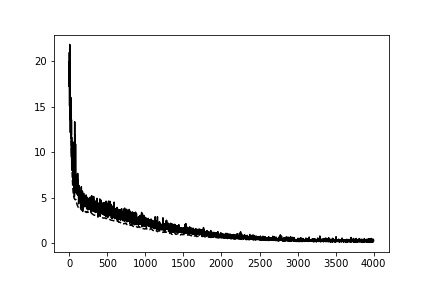}
    \caption*{CEGEN}
\end{minipage}
\begin{minipage}{0.33\textwidth}
    \includegraphics[width=\textwidth]{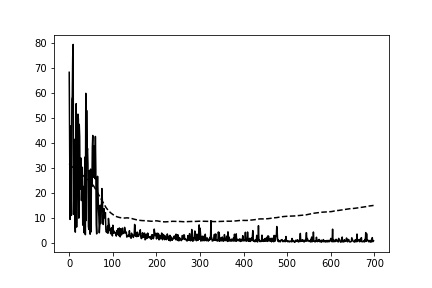}
    \caption*{TSGAN}
\end{minipage}
\begin{minipage}{0.33\textwidth}
    \includegraphics[width=\textwidth]{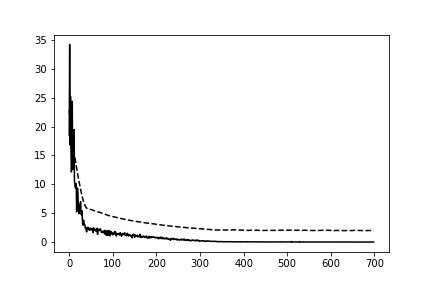}
    \caption*{COTGAN}
\end{minipage}
\begin{minipage}{0.32\textwidth}
    \includegraphics[width=\textwidth]{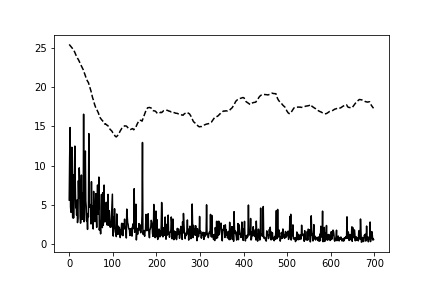}
    \caption*{SIGGAN}
\end{minipage}
\caption{Losses curves of the deep hedgers training to hedge a spread option between 3MAH Gas and 3MAH Coal. }
\label{fig:hedgerlossspread}
\end{figure}

\noindent Figure \ref{fig:hedgerlossspread} shows that the SIGGAN, COTGAN et TSGAN have difficulty converging correctly. In particular, the poor coverage performance of SIGGAN and TSGAN may be a consequence of their misrepresentation of historical correlations as shown in Table \ref{tab:correlation}.

\FloatBarrier

\begin{figure}[h]

\begin{minipage}{0.32\textwidth}
    \includegraphics[width=\textwidth]{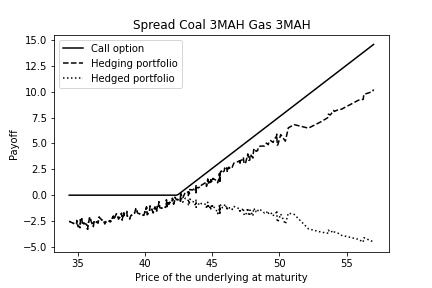}
    \caption*{Deep hedged Black-Scholes}
\end{minipage}
\begin{minipage}{0.32\textwidth}
    \includegraphics[width=\textwidth]{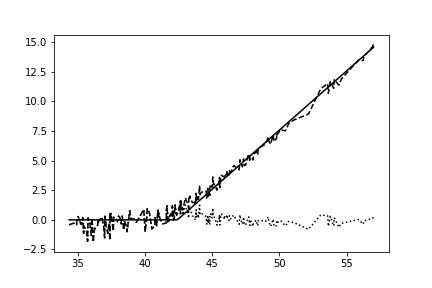}
    \caption*{CEGEN}
\end{minipage}
\begin{minipage}{0.33\textwidth}
    \includegraphics[width=\textwidth]{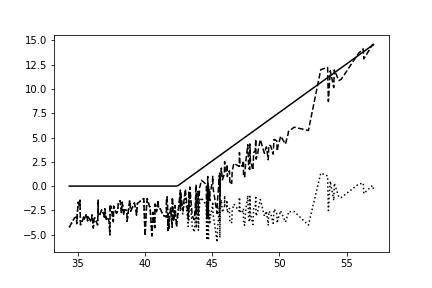}
    \caption*{TSGAN}
\end{minipage}
\begin{minipage}{0.33\textwidth}
    \includegraphics[width=\textwidth]{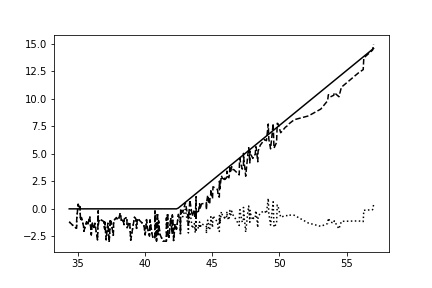}
    \caption*{COTGAN}
\end{minipage}
\begin{minipage}{0.33\textwidth}
    \includegraphics[width=\textwidth]{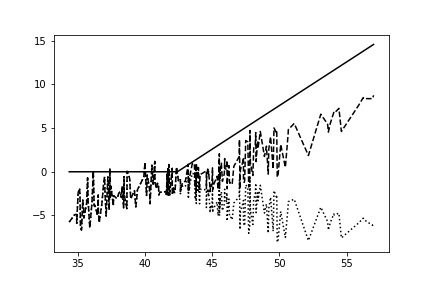}
    \caption*{SIGGAN}
\end{minipage}
\caption{Replicating portfolios of the spread call option.}
\label{fig:replgraphspread}
\end{figure}

\noindent These results are illustrated in Figure \ref{fig:replgraphspread}. Only CEGEN and COTGAN manage to reproduce the target broken line shape relatively well; but COTGAN's replicated portfolios are still noisy compared to CEGEN's.

\paragraph{Conclusion} Since the test case in this section is a mixture of the two previous test cases, it is not surprising that CEGEN still performs well in this situation. However, despite the very good performance in the delta hedging case, TSGAN fails to perform well in this situation. We explain this result by its poor representation of the correlation of the historical series as shown in the previous section and in the table \ref{tab:correlation}. 

\FloatBarrier
\section*{Conclusion}
A comparison between state-of-the-art deep generative methods is proposed on energy market applications. 
First, we evaluate the accuracy of the generations on energy commodity prices. 
Second, we present a study evaluating how these generators perform in a risk hedging task. Deep hedgers are trained on synthetic data produced by deep generators for option prices. CEGEN is the best performing generator in this situation. Its greatest strength lies in its replication of marginals and correlation and its consistency across different dataset. TSGAN is slightly less efficient on these aspects but is more generalist than CEGEN by not imposing Ito-type repesentation of time series. However, in the context of finance, it seems that a mixed Ito model/data driven model is more efficient than a purely data-driven model like TSGAN.

\FloatBarrier
\newpage
\appendix

\bibliography{biblio}


\end{document}